\documentclass[prd,floatfix,nofootinbib,superscriptaddress,showpacs,twocolumn,showkeys]{revtex4}
\usepackage{exscale}                  
\usepackage[intlimits]{amsmath}       
\usepackage{amsfonts}
\usepackage{amssymb,amscd}
\usepackage[dvips]{epsfig}                   
\usepackage{array}
\usepackage{wrapfig}
\usepackage{bbm}
\newcommand{\beqa}{\begin{eqnarray}}
\newcommand{\eeqa}{\end{eqnarray}}
\newcommand{\beq}{\begin{equation}}
\newcommand{\eeq}{\end{equation}}
\newcommand{\gS}[1]{#1\!\!\!\!\!\not~}	
\newcommand{\GS}[1]{#1\!\!\!\!\!\!\not~}

\newcommand{\pslash}{\gS{p}~}
\newcommand{\Pslash}{\GS{P}~}

\newcommand{\intq}{\int\!\!\frac{d^4q}{(2 \pi)^4}}

\newcommand{\tr}{\textrm{tr}}

\newcommand{\ONE}{\mathbbm{1}}

\newcommand{\lqcd}{\Lambda_{\mathrm{QCD}}}
\newcommand{\lqcdsq}{\Lambda^2_{\mathrm{QCD}}}
 
\begin{document}

\title{Beyond the rainbow: effects from pion back-coupling}
\author{Christian~S.~Fischer}
\affiliation{Institute for Nuclear Physics, 
 Darmstadt University of Technology, 
 Schlossgartenstra{\ss}e 9, 64289 Darmstadt, Germany}
\affiliation{Gesellschaft f\"ur Schwerionenforschung mbH, 
  Planckstr. 1  D-64291 Darmstadt, Germany.}
\author{Richard Williams}
\affiliation{Institute for Nuclear Physics, 
 Darmstadt University of Technology, 
 Schlossgartenstra{\ss}e 9, 64289 Darmstadt, Germany}

\date{\today}

\begin{abstract}
We investigate hadronic unquenching effects in light 
quarks and mesons. To this end we take into account the 
back-coupling of the pion onto the quark propagator within 
the non-perturbative continuum framework of 
Schwinger-Dyson equations (SDE) and Bethe-Salpeter equations (BSE). 
We improve on a previous approach by explicitly solving 
both the coupled system of DSEs and BSEs in the complex 
plane and the normalisation problem for Bethe-Salpeter 
kernels depending on the total momentum of the meson. 
As a result of our study we find considerable unquenching
effects in the spectrum of light pseudoscalar, vector and 
axial-vector mesons.
\end{abstract}


\maketitle

\section{Introduction}\label{sec:intro}

In hard scattering events mesons and baryons can be viewed as
bound states built up from partonic constituents, i.e. quarks and
gluons. This picture changes at low energies, where hadronic
effects play a more prominent r\^ole in the nonperturbative 
structure of hadrons. Of particular importance
are pion cloud effects which \emph{e.g.} have a direct impact on the
spin structure of the proton \cite{Thomas:2008bd}. 
Thus they need to be incorporated in bound-state calculations 
aiming at a realistic description of mesons and baryons.
 
For the spectrum of light pseudoscalar mesons, it is the axial-vector
Ward-Takahashi identity that governs the pattern of dynamical symmetry
breaking, rather than the intricacies of confinement or infrared
specifics of QCD. Within the framework of Schwinger-Dyson (SDE)
and Bethe-Salpeter (BSE) equations, this realisation has led to much
phenomenological success in describing pseudoscalar and
vector mesons~\cite{Maris:1997hd,Maris:1997tm,Maris:1999nt}, 
with bound-state masses, decay constants and electromagnetic form-factors
determined~\cite{Maris:2000sk,Bhagwat:2006pu} with surprising agreement to
experiment~\cite{Volmer:2000ek}. 

The success of this
description partly comes from the vector-vector coupling of the 
rainbow-ladder truncation, where the resultant spin-coupling to 
these $S$-wave mesons is dominant. For $P$-wave states however,
such as the scalar and axial-vector mesons, a richer tensor structure is
required from the quark-gluon vertex to properly account for the
observed spectrum. To this end, the scalar part of the quark-gluon
vertex, whose importance regarding confinement has been highlighted
in~\cite{Alkofer:2006gz}, may be the most relevant for introducing a 
spin-dependence to the interaction. We expect any beyond-the-rainbow 
study that generates such terms to have an effect on the spectrum of light
mesons.

However, there arise further complications in describing
the scalar and axial-vector sector. Not only are models employing the
rainbow-ladder truncation too attractive, yielding masses of
$800$-$900$~MeV~\cite{Maris:1999nt,Alkofer:2002bp}, but are these mesons 
even $q\bar{q}$ bound states? For the
scalars we have tetraquarks, multi-meson molecules, glueballs and the inevitable
hybrids as candidates (see e.g.
~\cite{Amsler:2004ps,Pennington:2005am}). Indeed, these exotics are not
restricted to the scalar sector, for the $a_1$ may be well described as a 
coupled-channel meson molecule~\cite{Wagner:2008gz}, amongst other equally plausible
descriptions \cite{Oller:1997ti,Oller:1998zr}. Clearly there is much work to be done in
providing a dynamical partonic description of these `mesons', not least
in understanding the fundamental interactions between quarks and gluons.

As a further step in this direction
we concentrate in this work on improving our model description of
$\bar{q}q$ bound-states. Based on an approximation scheme for the 
quark-gluon interaction
developed in \cite{Fischer:2007ze} and \cite{Fischer:2008sp} we study 
the pion back-reaction on the quarks and light mesons in the 
covariant Schwinger-Dyson/Bethe-Salpeter
approach to Landau gauge QCD \cite{Maris:2003vk,Fischer:2006ub}. As a 
consequence
we take into account pion cloud effects in the light meson spectrum.
To this end we solve the coupled system of quark Schwinger-Dyson equations
and Bethe-Salpeter equations for 
pseudoscalar, vector and axial-vector mesons in the complex 
plane. This provides for an important technical progress compared
to the calculations in Ref.~\cite{Fischer:2007ze}, where the
so-called real-value approximation has been used. As a further
technical complication we then have to solve the normalisation
problem for Bethe-Salpeter amplitudes generated by a kernel
depending on the total momentum of the meson. 

The paper is organized as follows. In Sec.~\ref{sec2}. we
summarise our approximation scheme for the quark-gluon interaction
developed in detail in \cite{Fischer:2007ze,Fischer:2008sp}. We specify our
model interaction and discuss the technical complications
that appear in the normalisation condition for the bound-state
amplitudes. We explicitly verify that our interaction satisfies
the axial-vector Ward-Takahashi identity. 
In Sec.~\ref{sec3}. we then provide the details of our 
calculation and present our results. We conclude with an outlook
in Sec.~\ref{sec4}. Some technical discussions are relegated to an appendix.

\section{The quark-gluon interaction}\label{sec2}

\subsection{General truncation}

In the past, a widely used practical approximation scheme for 
the coupled system of quark-SDE and meson-BSE is the 
rainbow-ladder approximation. Its most important property is
that it satisfies the non-singlet axial-vector Ward-Takahashi
identity (axWTI) which guarantees that the pion in the chiral limit is both
a Goldstone boson and a bound state of massive constituents.
This scheme has a history of remarkable successes (summarised 
{\it e.g.} in~\cite{Maris:2003vk}). However, there are also 
shortcomings that limit the credibility of such an approximation. 
Consequently, several efforts have been made to extend this scheme, 
see {\it e.g.}~\cite{Bender:1996bb,Watson:2004kd,Bhagwat:2004hn,
Fischer:2005en,Maris:2005tt,Matevosyan:2006bk}. In the following, we will continue
the strategy employed in \cite{Fischer:2007ze} to extend the 
rainbow-ladder scheme by taking into account pion cloud effects
in the quark-gluon interaction. These effects can be accounted
for by a particular class of interaction diagrams that only appear
in the unquenched theory. Our strategy is complementary to the one 
followed in \cite{Watson:2004jq}, where a different class of unquenching 
effects have been investigated. The aim there was to provide for
hadronic intermediate states in bound-state calculations which 
generate a finite width of meson spectral functions. 
It is also complementary to the investigations reported 
in~\cite{Fischer:2003rp,Fischer:2005en}, where unquenching effects 
in the gluon polarization have been considered.
 
\begin{figure}[ht]
\centerline{\includegraphics[width=0.9\columnwidth]{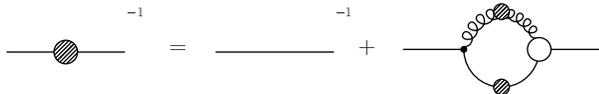}}
\caption{The Schwinger-Dyson equation for the fully dressed
quark propagator}\label{fig:dse1}
\end{figure}

We start with the SDE for the quark propagator, given
diagrammatically in Fig.~\ref{fig:dse1}. 
With the fully dressed
inverse quark propagator $S^{-1}(p)=i \pslash A(p^2) + B(p^2)$ and
its bare counterpart $S^{-1}_{0}(p) = i \pslash + m$, the equation
is given by
\begin{eqnarray}
  S^{-1}(p) &=& Z_{2}S^{-1}_{0}(p)\nonumber\\[-2mm] \label{DSE1}\\[-2mm]
  +&&\hspace{-15pt} g^{2}C_{F}Z_{1F}\intq\,  \gamma_{\mu}S(q)\Gamma_{\nu}(q,k)
  D_{\mu\nu}(k)   \,,\nonumber
\end{eqnarray}
with $k=p-q$, the Casimir $C_{F}=(N_{c}^{2}-1)/(2N_{c})$ and 
the renormalization factors $Z_{1F}=Z_2/\widetilde{Z}_3$ of the 
quark-gluon vertex, $Z_2$ of the quark propagator and $\widetilde{Z}_3$
of the ghost dressing function. The quark dressing functions $A(p^2)$ and 
$B(p^2)$ can be recombined into the quark mass 
$M(p^2) = B(p^2)/A(p^2)$ and the quark wave function 
$Z_f(p^2) = 1/A(p^2)$. The quark self-energy depends on the fully 
dressed quark-gluon vertex $\Gamma_{\nu}(q,k)$ and the gluon propagator 
\begin{eqnarray}
D_{\mu\nu}(k) = \left(\delta_{\mu \nu} 
- \frac{k_\mu k_\nu}{k^2}\right)                                        
\frac{Z(k^2)}{k^2}\,, \label{gluon}
\end{eqnarray}
with the gluon dressing function $Z(k^2)$. Throughout this paper 
we will work in the Landau gauge, which guarantees the 
transversality of (\ref{gluon}). The explicit expression for $Z(k^2)$
used in this work will be given in Sec.~\ref{model}.

\begin{figure}[t]
\centerline{\includegraphics*[width=0.98\columnwidth]{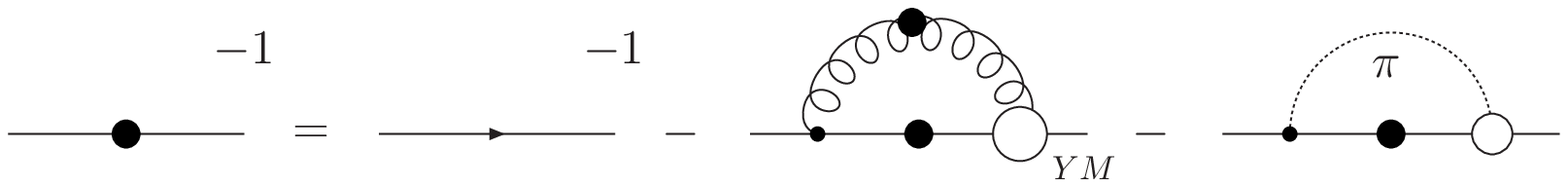}}
\caption{The approximated Schwinger-Dyson equation for the 
quark propagator with effective one-gluon exchange and 
one-pion exchange. 
\label{fig:quarkdse2}}
\vspace*{10mm}
\centerline{\includegraphics*[width=0.98\columnwidth]{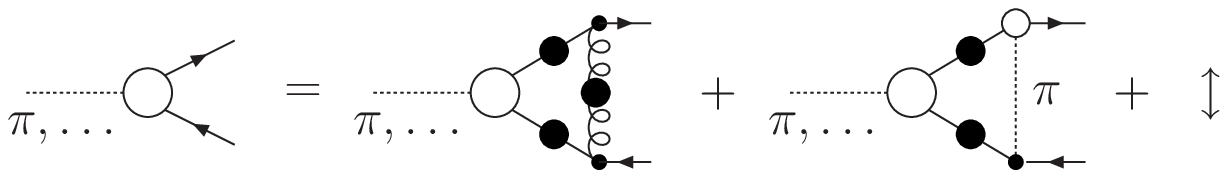}}
\caption{The Bethe-Salpeter equation corresponding to the 
quark self-energy of Fig.~\ref{fig:quarkdse2}. The up-down arrow indicates 
an averaging procedure of the pion-exchange diagram with its counterpart
where the upper pion-quark vertex is bare and the lower one dressed.
\label{fig:BSE}}
\vspace*{5mm}
\end{figure}

The key quantity in the SDE for the quark propagator is the quark-gluon
vertex. It satisfies its own Schwinger-Dyson equation which contains
various basically unknown four-point functions. The structure of the
Yang-Mills part of this equation and the resulting pattern of 
dynamical chiral symmetry breaking has been discussed recently in 
detail in Ref.~\cite{Alkofer:2006gz}. Here we are interested 
in the hadronic contributions to this vertex. These have been discussed in 
Ref.~\cite{Fischer:2007ze} and \cite{Fischer:2008sp}, where two slightly
different approximations in terms of one-pion exchange have been constructed.
Whereas the scheme of \cite{Fischer:2007ze} resulted in a very strong 
back-reaction of the pion onto the quark propagator in disagreement with
corresponding lattice results, the modified scheme in 
\cite{Fischer:2008sp} seems to represent these effects quantitatively 
correctly. We therefore resort to the latter scheme here, whose 
diagrammatic expressions for the quark-SDE and the corresponding 
Bethe-Salpeter equation for quark-antiquark bound states are given in 
Figs.~\ref{fig:quarkdse2} and \ref{fig:BSE}. 

The effective self-interaction of the quark is thus given by a pure
Yang-Mills part and a hadronic part approximated by one-pion exchange.
The dressed coupling of the pion to the quark line is hereby described by its
Bethe-Salpeter vertex function, whereas the bare one is given by 
$\tau^j \gamma_5 Z_2$ with the generators $\tau^j$ of flavour SU(2). 

Certainly it is important that the explicit form of both the Yang-Mills 
part and the pionic part of the
interaction is such that the axial-vector Ward-Takahashi identity
is satisfied. Naturally, as a first guess for the Yang-Mills part one
takes a rainbow-ladder form of the interaction, \emph{i.e.} one chooses a 
model for the fully dressed Yang-Mills part of the quark-gluon 
vertex $\Gamma_\nu^{\rm YM}$ of the form 
\begin{equation}
\Gamma_\nu^{YM}(p_1,p_2,p_3) \;=\; \gamma_\nu\, 
{Z_2}/{\widetilde{Z}_3} \, \, \Gamma^{YM}(p_3^2) \,, \label{v1}
\end{equation}
where we denote the quark momenta by $p_1$ and $p_2$ and the gluon momentum by $p_3$. 
The explicit expression for $\Gamma^{\rm YM}(p_3^2)$ 
is detailed below. A similar form has been used in~\cite{Bhagwat:2003vw,Fischer:2005nf}, 
where quenched lattice results for the quark propagator have been
reproduced. Note, however, that (\ref{v1}) involves only the
$\gamma_\nu$-part of the tensor structure of the full Yang-Mills part
of the vertex. It has been shown in the analysis of the full quark-gluon 
vertex of Ref.~\cite{Alkofer:2006gz} that such a model cannot capture 
all essentials of dynamical chiral symmetry breaking. Indeed, from the
results of this work we will draw the very same conclusion from 
our calculation of the light meson spectrum.

In general one can decompose the quark-gluon vertex $\Gamma^\nu(p_1,p_2,p_3)$ 
into 12 different tensor structures, given by
\begin{equation}
\Gamma_\nu(p_1,p_2,p_3)=\sum_{i=1}^{12} \lambda_i(p_1,p_2,p_3) L^i_\nu (p_1,p_2,p_3)\,.
\end{equation}
The details of this basis are given in Ref.~\cite{Skullerud:2002ge}. 
Here, we only note that $L^1_\nu = \gamma_{\nu}$; the explicit forms 
of the other structures are not needed. In principle, all twelve 
tensor structures are important in the infrared and intermediate 
momentum regime \cite{Skullerud:2003qu,Alkofer:2006gz}. We have 
verified by projection methods that the pion-exchange diagram 
in the vertex-SDE indeed contributes to all these structures. 
Therefore the resulting one-pion contribution in the propagator SDE
Fig.~\ref{fig:quarkdse2} generates unquenching effects in all components
of the quark-gluon interaction. 

An explicit expression for the one pion-exchange kernel that satisfies
the axial-vector Ward-Takahashi identity has been constructed in 
\cite{Fischer:2007ze}. Here we give a similar construction which takes
into account the modifications suggested in \cite{Fischer:2008sp}.
The resulting SDE for the quark propagator is given by
\begin{widetext}
\begin{eqnarray}
  S^{-1}(p) = Z_2 \,S^{-1}_0(p) &+& g^2 \, C_F \, (Z_{2})^2 
  \int \frac{d^4q}{(2\pi)^4} \, \gamma_\mu \, S(q) \, \gamma_\nu \, 
  \left(\delta_{\mu\nu}-\frac{k_\mu k_\nu}{k^2}\right) \frac{Z(k^2) 
    \Gamma_{\rm YM}(k^2)}{k^2} \nonumber\\
  &-&3 \int \frac{d^4q}{(2\pi)^4} \left[      
    Z_2 \gamma_5 \, S(q)\,
    \Gamma_{\pi}\left(\frac{p+q}{2};q-p\right) + 
    Z_2 \gamma_5 \, S(q)\,
    \Gamma_{\pi}\left(\frac{p+q}{2};p-q\right)
    \right] \frac{D_{\pi}(k^2)}{2} \label{quark}
\end{eqnarray}
\end{widetext}
with $k=p-q$ and $D_{\pi}(k)=(k^{2}+M_{\pi}^{2})^{-1}$ being the pion
propagator.
The factor $3$ in front of the pion contribution stems from the 
flavour trace and represents contributions from $\pi_+, \pi_-$ and $\pi_0$. 
These are treated on equal footing in the isospin-symmetric limit
adopted here.

The corresponding expression for the Bethe-Salpeter equation for light mesons reads
\begin{widetext}
\begin{eqnarray}
\Gamma_{tu}^{(\mu)}(p;P)\!\!&=& \!\!\!\!\int \frac{d^4k}{(2\pi)^4}
\left\{K_{tu;rs}^{\textrm{YM}}(p,k;P) 
                     + K_{tu;rs}^{\textrm{pion}}(p,k;P)\right\}
			   \left[S(k_+)\Gamma^{(\mu)}(k;P)S(k_-)\right]_{sr}
\label{eq:bse2}
\end{eqnarray}
with the kernels $K_{tu;rs}^{\textrm{YM}}$ and $K_{tu;rs}^{\textrm{pion}}$ given by
\begin{eqnarray}  
 K_{tu;sr}^{\textrm{YM}}(q,p;P)
  &=&
  \frac{g^2 \, Z(k^2)\,  \Gamma^{\textrm{YM}}(k^2) \, Z_{1F}}{k^{2}}
  \left(\delta_{\mu\nu}-\frac{k_{\mu}k_{\nu}}{k^{2}}\right)
  \left[\frac{\lambda^{a}}{2}\gamma_{\mu}\right]_{ts}
  \left[\frac{\lambda^{a}}{2}\gamma_{\nu}\right]_{ru}\,, \label{YMkernel}\\
  K_{tu;rs}^{\textrm{pion}}(q,p;P)
  &=&
  \frac{1}{4}
      [\Gamma^j_{\pi}]_{ru}\left(\frac{p+q-P}{2};p-q\right)
      [Z2 \tau^j \gamma_5]_{ts}
       D_{\pi}(p-q)
  \nonumber\\
  &+&\frac{1}{4}
      [\Gamma^j_{\pi}]_{ru}\left(\frac{p+q-P}{2};q-p\right)
      [Z2 \tau^j \gamma_5]_{ts}
      D_{\pi}(p-q)    
  \nonumber\\
  &+&\frac{1}{4}
      [\Gamma^j_{\pi}]_{ts}\left( \frac{p+q+P}{2};p-q\right)
      [Z2 \tau^j \gamma_5]_{ru}
      D_{\pi}(p-q)    
  \nonumber\\
  &+&\frac{1}{4}
      [\Gamma^j_{\pi}]_{ts}\left( \frac{p+q+P}{2};q-p\right)
      [Z2 \tau^j \gamma_5]_{ru}
      D_{\pi}(p-q)    
      \,.\phantom{ccc}\label{Kapprox1}
\end{eqnarray}
\end{widetext}
Here $\Gamma^{(\mu)}(p;P)$ is the Bethe-Salpeter vertex function of a
quark-antiquark bound state specified below, whereas $\Gamma^j_{\pi}$
with flavour index $j$ is the corresponding vertex function of a pion. 
The momenta
$k_+=k+P/2$ and $k_-=k-P/2$ are such that the total momentum $P$ 
of the meson is given by $P=k_+-k_-$ and the relative momentum
$k=(k_++k_-)/2$. The Latin indices ($t,u,r,s$) of the
kernels refer to colour, flavour and Dirac structure.

The Bethe-Salpeter vertex function $\Gamma(p;P)^{(\mu)}$ can be 
decomposed into at most eight Lorentz and Dirac structures.  The structure is 
constrained by the transformation properties under CPT of the meson
we wish to describe~\cite{LlewellynSmith:1969az}. In particular, our pseudoscalar, scalar and vector
have quantum numbers $J^{P}$ of $0^-$, $0^+$ and $1^-$, respectively. The
axial-vector can be parameterised in two different ways, dependent on
its transformation under charge conjugation,
$J^{PC}=1^{++}$ and $1^{+-}$.

\begin{table}[t]
\begin{eqnarray}
\begin{array}{@{}lcc|cc}
                       & & \textrm{Component}      & 0^- & 0^+ \\
\hline
T_1(p;P) &:&  \ONE                  & \circ  & \circ \\
T_2(p;P) &:&  -i\Pslash 		& \circ  & \bullet\\
T_3(p;P) &:&  -i\pslash			& \bullet& \circ\\
T_4(p;P) &:& \left[ \pslash,\Pslash
\right]					& \circ  & \circ\\
\end{array}\nonumber
\end{eqnarray}
\caption{The four Dirac structures describing a meson of spin $J=0$.
  For the pseudoscalar, these have an associated factor of $\gamma_5$ to
  account for parity. Circles indicate that the component is part of the 
  basis decomposition of the meson, with filled circles indicating that 
  the component is multiplied by $\left(p\cdot P\right)$. This ensures 
  that for equal-mass constituents a Chebyshev decomposition of even 
  order need only been employed for the scalar amplitudes $F_i$.}\label{basis0}
\end{table}
\begin{table}[t]
\begin{eqnarray}
\begin{array}{@{}lcc|c|cc}
         & & \textrm{Component}      & 1^-  & 1^{++} & 1^{+-} \\
\hline
T_1^\mu(p;P) &:& i\gamma_T^\mu      & \circ   & \circ &  \\[1mm]
T_2^\mu(p;P) &:& \gamma^\mu_T\Pslash& \circ   &\bullet&  \\[1mm]
T_3^\mu(p;P) &:& -\gamma^\mu_T\pslash +p^\mu_T\ONE&\bullet &\circ&
 \\[1mm]
T_4^\mu(p;P) &:& i\gamma^\mu_T\left[ \Pslash,\pslash
\right]+2ip^\mu_T\Pslash          &\circ&\circ& \\[1mm]
T_5^\mu(p;P) &:& p^\mu_T\ONE      &\circ   &     &\circ \\[1mm]
T_6^\mu(p;P) &:& i p^\mu_T\Pslash &\bullet &     & \circ\\[1mm]
T_7^\mu(p;P) &:& -i p^\mu_T\pslash&\circ   &       & \bullet \\[1mm]
T_8^\mu(p;P) &:& p^\mu_T\left[ \Pslash,\pslash \right]&\circ &     & \circ \\
\end{array}\nonumber
\end{eqnarray}
\caption{The eight Dirac structures describing a meson of spin $J=1$.
For the axial-vectors, there is an additional factor of $\gamma_5$ to
account for parity. Circles indicate that the component is included in 
the basis decomposition of the meson, with filled circles indicating 
the component is multiplied by $\left( p\cdot P\right)$. This ensures 
that an even Chebyshev decomposition is sufficient for equal-mass 
constituents. The subscript $T$ indicates transversality with respect 
to the total momentum.}\label{basis1}
\end{table}
Taking this into consideration, we can write
down a general basis for each desired meson amplitude,
\begin{equation}
\Gamma_M^{(\mu)}(p;P)=\Bigg\{\begin{array}{cc}
\sum_{i} \phantom{\gamma_5}F_i(p;P)T^{(\mu)}_i(p;P) &\,\,\, J^{P} = 0^+, 1^-\\
\sum_{i} \gamma_5F_i(p;P)T^{(\mu)}_i(p;P) &\,\,\, J^{P}=0^-, 1^+\\
\end{array}
\end{equation}
where the components $T_i^{(\mu)}$ are given in Tables~\ref{basis0} and
\ref{basis1}.

In particular, the pion is
given by the following form
\begin{eqnarray}
\Gamma_\pi(p;P)\!\!&=&\!\!\!\! \gamma_{5}\Big[F_1(p;P)
-i\Pslash F_2(p;P)\nonumber\\[-2mm]\label{pion}\\[-2mm]
&&\hspace{-9mm}-i\pslash \left(p\cdot P\right)F_3(p;P)
-\left[\Pslash,\pslash\right]F_4(p;P)\Big]\, .\nonumber
\end{eqnarray}

\subsection{Model details \label{model}}
Next we need to specify the details of our model interaction. For
the Yang-Mills part of the interaction we choose two different
model ans\"atze that have been employed already in previous works.
Here, the gluon dressing function $Z(k^2)$ from (\ref{gluon})
and the Yang-Mills part $\Gamma^{\rm YM}(k^2)$ of the quark-gluon vertex 
Eq.~(\ref{v1}) are combined and represented by a single function.
For the Maris-Tandy (MT) model \cite{Maris:1999nt} this function 
is given by 
\begin{eqnarray}
Z(k^2) \Gamma^{\textrm{YM}}(k^2) &=&  \frac{4\pi}{g^2} 
       \bigg( \frac{\pi}{\omega^6}D q^4 \exp(-q^2/\omega^2)\nonumber\\
	&+&\frac{2\pi \gamma_m}{\log\left( \tau+\left(1+q^2/\lqcdsq \right)^2\right)}
	\nonumber\\[0.mm]
	&&\times \left[ 1-\exp\left(-q^2/\left[ 4m_t^2 \right]\right) \right]\bigg)\;, \label{model1}
\end{eqnarray}
with 
\begin{equation}
\begin{array}{lcl}
m_t= 0.5\;{\rm GeV}\,&,&\qquad \tau\;=\;\mathrm{e}^2-1\,\\
 \gamma_m=12/(33-2N_f)\,&,&\quad \lqcd\;=\;0.234\,{\rm GeV}\, .
 \end{array}\nonumber
\end{equation}
This model has been employed successfully in various calculations of
meson spectra, form factors and decay constant within a rainbow-ladder 
truncation; see \cite{Maris:2003vk,Maris:2005tt} for overviews of the 
results. The remaining parameters $D$, $\omega$ and the quark mass are
fitted to pion observables, and detailed in Sec.~\ref{sec3}.

The second model we employ for the Yang-Mills part of the quark-gluon
interaction represents the infrared structure of both the gluon dressing
function $Z(k^2)$ and the quark-gluon vertex $\Gamma^{\rm YM}(k^2)$, in 
quenched approximation (up to effects due to hairpin diagrams)
as calculated in Refs.~\cite{Alkofer:2003jj,Alkofer:2006gz}. 
It has been used to show that infrared divergences in vertex functions 
can account for the topological mass of the $\eta'$-meson 
\cite{Alkofer:2008et}, and is given by
\begin{eqnarray}
Z\left(k^2\right) &=& \left(\frac{k^2}{k^2+\Lambda^2_{\mathrm{QCD}}}\right)^{2\kappa}
                  \left(\frac{\alpha_{\mathrm{
			fit}}\left(k^2\right)}{\alpha_\mu}\right)^{-\gamma}\;,\nonumber\\
\Gamma^{\mathrm{YM}}(k^2) &=& \left(\frac{k^2}{k^2+
d_2}\right)^{-1/2-\kappa}\label{model2}\\
&&\hspace{-1.5cm}\times\Bigg( \frac{ d_1}{d_2+k^2}+
\frac{ k^2 d_3}{d_2^2+\left(k^2- d_2\right)^2}
+\frac{k^2}{ { \lqcdsq}+k^2}
\nonumber
\\[-1.2mm]
&&\hspace{-1.5cm}\times\left[\frac{4\pi}{\beta_0\alpha_\mu}
\left(\frac{1}{\log\left(\frac{k^2}{\lqcdsq}\right)}-\frac{\lqcdsq}{k^2-\lqcdsq}\right)\right]^{-2\delta}
	    \Bigg)\nonumber
\end{eqnarray}
with the gluon momentum $k^2$, the one-loop value 
$\gamma = (-13 N_c + 4 N_f)/(22 N_c - 4 N_f)$ for the anomalous dimension of the 
re-summed gluon propagator, the corresponding one $\delta = \frac{-9 N_c}{44 N_c - 8 N_f}$
for the ghost dressing function and $\alpha_\mu=0.2$ at the renormalisation scale 
$\mu^2=170$GeV$^2$. We use $\lqcdsq=0.52$ GeV$^2$ similar to the scale 
obtained in Ref.~\cite{Alkofer:2003jj}. The infrared exponent $\kappa$ has been 
determined analytically in \cite{Zwanziger:2001kw,Lerche:2002ep} and is given by
$\kappa=\left(93-\sqrt{1201}\right)/98\simeq0.595$. Note that this form
is slightly modified to that reported
in~\cite{Alkofer:2008et,Fischer:2008sp}, allowing more freedom in
separating the scales in the quark-gluon vertex and the gluon
propagator. 
The parameter $d_1$ sets the strength of the interaction in the infrared.  The 
parameters $d_2$ and $d_3$ determine the location and size of a term added to 
give additional integrated strength in the intermediate momentum
regions. These three parameters, together with the current quark mass,
are constrained by pion observables and the generation of a topological
charge in the pseudoscalar isosinglet channel.
The running coupling from the ghost-gluon vertex, 
$\alpha_{\rm fit}(p^2)$, is parameterised such that the numerical results
for Euclidean scales are accurately reproduced \cite{Alkofer:2003jj}:
\begin{eqnarray}
\alpha_{\rm fit}(p^2) &=&  \frac{\alpha_s(0)}{1+p^2/\Lambda^2_{\rm QCD}}
+ \frac{4 \pi}{\beta_0} \frac{p^2}{\lqcdsq+p^2}\nonumber\\[-2mm]\label{eqn:alpha}\\[-2mm]
&& \left(\frac{1}{\ln(p^2/\lqcdsq)}
- \frac{1}{p^2/\lqcdsq -1}\right)\;.\nonumber 
\end{eqnarray}
Here $\beta_0=(11N_c-2N_f)/3$, and $\alpha_S(0)$ is the fixed point in the
infrared, calculated to be $\alpha_S(0)=8.915/N_c$ for our choice of
$\kappa$. In this work, we will take $N_f=2$. 

\begin{figure}[t]
\centerline{\includegraphics[width=0.8\columnwidth]{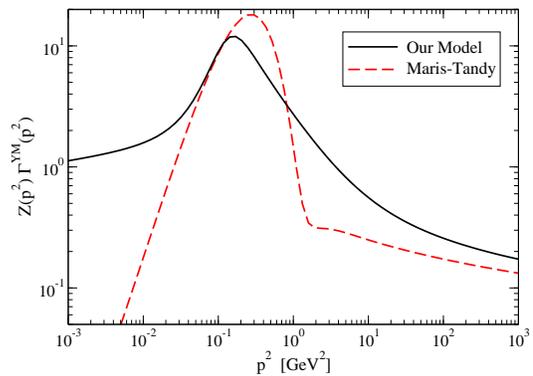}}
\caption{A log-log plot of the effective quark-gluon interaction
$Z \Gamma^{YM}$ for the two models we employ.
Differences in the UV behaviour are attributed to the different scale
$\lqcdsq$ in use. Note both exhibit enhancement in the intermediate
momentum region.}\label{fig:coupling}
\end{figure}

The resulting two effective interactions from the Maris-Tandy model (\ref{model1})
and our soft-divergence model (\ref{model2}) are plotted against each other in
Fig.~\ref{fig:coupling}. The ultraviolet running of both interactions are the
same except for differences in the scale employed. Whereas the Maris-Tandy model
used a scale corresponding to the $\overline{MS}$ scheme, we are working in a momentum
subtraction scheme where the scale is larger by more than a factor of 3. 
Both interactions share an enhancement in the intermediate
momentum region which provides the necessary interaction strength for dynamical chiral
symmetry breaking. The change of slope in the infrared region of our interaction
is related to the topological mass component of the $\eta'$ in the chiral limit
\cite{Alkofer:2008et}.

As for the pion part of our interaction, we approximate the full pion Bethe-Salpeter
wave function (\ref{pion}) in the quark-SDE (\ref{quark}) and the kernel
of the BSE (\ref{Kapprox1}) by the leading amplitude in the chiral limit given
by \cite{Maris:1997hd}
\begin{equation}
\Gamma^j_{\pi}(p;P) = \tau^j \gamma_5 \frac{B(p^2)}{f_\pi}\;. \label{piapprox}
\end{equation}
Here $B(p^2)$ is the scalar dressing function of the quark propagator
in the chiral limit.
This approximation omits the back-coupling effects of the three sub-leading 
amplitudes. From the calculation of ref.~\cite{Fischer:2007ze}, where full 
back-coupling has been evaluated in a real value approximation, we expect this 
omission to lead to an error of only a few percent as concerns meson masses and
of about 10-20\% as concerns decay constants. Considering the exploratory
character of our work this seems certainly tolerable. The replacement of the
leading physical pion amplitude $F_1(p;P)$ at $m_\pi=138$ MeV by the 
(normalised) chiral limit value $F_1(p;P) \approx \frac{B(p^2)}{f_\pi}$ is 
also correct on the few percent level as demonstrated in the appendix. 
Note that we use this approximation only for the internal pion which mediates the
interaction. The great advantage
of the approximation (\ref{piapprox}) compared to the full back-coupling performed
in Ref.~\cite{Fischer:2007ze} is that we can then fully take into account the 
quark propagator in the complex plane as necessary input into the Bethe-Salpeter
equation.

\subsection{Decay constants and normalisation \label{res}}

The Bethe-Salpeter amplitudes for the pseudoscalar mesons can be used 
to obtain the corresponding weak decay constants $f_{\pi}$, which will prove 
useful in fitting parameters to observables. In order to calculate these one must 
firstly normalise the amplitudes.  The normalisation condition is derived by demanding 
that the residue of the pole in the four-point quark-antiquark Green's function 
(from which the BSE is derived) be unity \cite{Tandy:1997qf,Nakanishi:1969ph}.  
Using conventions such that $f_\pi=93$ MeV, and for momenta partitioned
equally between the two constituents, it reads
\begin{eqnarray}\label{norm}
\delta^{ij}&=&2\frac{\partial}{\partial P^2} \tr\int
\frac{d^4k}{(2\pi)^4}\\
& \Bigg[&\!\!\!\!3 \, \, \bigg(
\overline{\Gamma}_\pi^i(k,-Q)  S(k+P/2)\Gamma_\pi^j(k,Q)S(k-P/2)
\bigg) \nonumber\\
\!\!&+&\!\!\!\!\int \frac{d^4q}{(2\pi)^4} 
[\overline{\chi}_\pi^i]_{sr}(q,-Q)
K_{tu;rs}^{\textrm{pion}}(q,k;P)[{\chi}_\pi^j]_{ut}(k,Q)\Bigg]\;,\nonumber
\end{eqnarray}
where $Q^2=-M^2$ is fixed to the on-shell meson mass,
the trace is over Dirac matrices and the 
Bethe-Salpeter wave-function $\chi$ is defined by
\begin{equation}
\chi_\pi^j(k;P) = S(k+P/2)\Gamma^j_\pi(k,P)S(k-P/2)\,.
\end{equation}
The conjugate vertex function $\bar{\Gamma}$ is given by
\begin{equation}
\bar{\Gamma}(p,-P)=C \Gamma^T(-p,-P)C^{-1}\;,
\end{equation}
with the charge conjugation matrix $C=-\gamma_2\gamma_4$.
The leptonic decay constant characterising the pion coupling 
to the point axial field is subsequently given by \cite{Tandy:1997qf}
\begin{equation}
f_{\pi}\!=\!Z_2\frac{3}{M^2} \tr \!\!\int\!\frac{d^4k}{(2\pi)^4}
\Gamma_\pi(k,-P) \,S(k_+) \,\gamma_5 \,\Pslash \,S(k_-)\;,
\label{eq:fpi}
\end{equation}
where again the trace is over Dirac matrices, and $k_+=k+P/2$,
$k_-=k-P/2$.   There exist analogous
expressions for the vector mesons~\cite{Maris:1999nt}, where an additional 
factor of $1/3$ must be included due to summation over the polarisation tensor.

One may write a similar equation to (\ref{eq:fpi}), which corresponds to
the residue of the pseudoscalar vertex:
\begin{equation}
r_{\pi}\!=Z_2 Z_m\, 3\, \tr \!\!\int\!\frac{d^4k}{(2\pi)^4}
\Gamma_\pi(k,-P) \,S(k_+) \,\gamma_5 \,S(k_-)\;.
\label{eq:rpi}
\end{equation}
The axWTI imposes a relationship between these two residues, known as
the generalised Gell-Mann--Oakes--Renner relation~\cite{Maris:1997tm},
which must hold at and beyond the chiral limit:
\begin{equation}
f_\pi m_\pi^2 = r_\pi\left( m_u(\mu^2) + m_d(\mu^2) \right)\;,
\end{equation}
where $m_u$, $m_d$ are the masses of the up and down quarks at the renormalisation
point $\mu^2$, and in this work considered to be degenerate. Confirming
this relation serves as a check of both our numerics and indicates how
well our kernel satisfies the axWTI. We find excellent agreement with
errors smaller than $1\%$ for pion masses up to $m_\pi \simeq 600$ MeV.

\section{Numerical methods and results \label{sec3}} 

\subsection{Solving the BSE}
We solve the BSE for the meson amplitude $\Gamma^{(\mu)}(p;P)$
via matrix methods using the following procedure. First, we project
our BSE onto the scalar amplitudes of our meson decomposition.
This gives rise to either four or eight coupled integral equations for
the $F_i(p;P)$. To make manifest the angular dependence of the
amplitude functions, we treat the total momentum $P^2$ as a parameter
and expand the function as a series of Chebyshev polynomials in the
angle
$\widehat{p\cdot P} = p\cdot P/|pP|$
\begin{equation}
F_i(p;P) = \sum_k (i)^k
F^k_i(p^2;P^2)T_k(\widehat{p\cdot P})\;.
\end{equation}
The functions $F_i^k(p^2;P^2)$ are projected out through use of the
orthonormal properties of the Chebyshev polynomials. With the angular
dependence made explicit, we can evaluate numerically the two
non-trivial angles appearing in the integration measure. We cast the
remaining radial integral in the form of a matrix equation by matching
the external momenta to the radial loop momenta, $p^2_j = k^2_j$ at the
abscissae of our integration nodes. Thus
the amplitude of our BS equation is projected onto the decomposition
$F^k_i(p^2_j;P^2)$. Schematically we are solving
\begin{equation}
\Gamma = \lambda\, {\bf K}\cdot \Gamma\;,
\end{equation}
for the column vector $\Gamma$ as a parametric equation in $P^2=-M^2$
with $M$ the mass of the meson. A bound-state corresponds to solutions with $\lambda=1$, which
as an eigenvalue problem is equivalent to satisfying the condition
$\det \left( \ONE-{\bf K} \right)=0$. Because of the tractable nature of our
unquenching prescription, the computational effort required to determine
the Bethe-Salpeter amplitude increases from a few seconds to a few
minutes. However, we see in the next section that substantial effort is
required to calculate the normalisation condition with any kernel
involving momentum exchange of the meson.

\subsection{Calculating the normalisation}
The first term of (\ref{norm}), independent of the kernel, is easily evaluated since
the quark propagator derivatives are directly calculable in our
approach; either by applying the derivative to the quark-SDE, and
solving
the resulting integral equation, or by finite difference methods. The
second term is substantially more complicated for two reasons. First,
we must determine the derivative of our exchange kernel with respect to
the total momentum dependence -- a tractable but algebraically complicated
task. The second problem is the double integration over two
four-momenta, $k$ and $q$. Decomposing the measure into hyperspherical
coordinates, where none of the angles may be evaluated
trivially, results in the need to evaluate an eight dimensional
integral numerically. Resigning ourselves to this computationally
intensive task, we evaluate the
integral of (\ref{norm}) for $P^2 = Q^2\pm 2^n\Delta$, and employ the appropriate
Richardson improved centered difference formula. Typically, we use a
four-point rule for which $n=0,1$ though a two-point rule is sufficient,
provided we choose $\Delta$ optimally.
Unfortunately each double integral must still be evaluated, which we
tackled using both Monte-Carlo and Gaussian quadrature techniques. For the
latter, a small-scale computing cluster is mandatory for even moderate
grid sizes. Monte-Carlo performs only marginally better in terms of
necessary computing time, since the
statistical errors must be minimised such that the numerical derivative
is meaningful. With due care, however, both methods obviously yield the
same result. 

In the next section, we give results for the leptonic decay
constants for the pion and rho with and without consideration of the
normalisation double-integral. For a physical pion, the difference is
only about $3\%$ and smaller still for the rho. However, as we reduce
the pion mass and approach the chiral limit the necessity of including
the full normalisation condition becomes readily apparent.

\subsection{Results}
The parameters of the two models we employed -- that of Maris-Tandy and
our model -- were tuned such that pion observables were well
reproduced with the inclusion of the hadronic exchange kernel. For our
interaction, which is capable of generating a topological charge, the
eta/eta' splitting provides an additional constraint. 

For the
Maris-Tandy model, we find a unique choice of $\omega$ and $D$ for a
given quark mass by fitting to $m_\pi$ and $f_\pi$. We attempted to
maximise the rho meson mass, though we did not exhaust
the available parameter space due to the complexity in evaluating the
full normalisation condition. Consequently we employ $\omega=0.37$~GeV 
and $D=1.45$~GeV$^2$, with a quark mass of $3.7$~MeV at $\mu=19$~GeV.

The parameters of our model interaction were fit in a similar fashion,
with the additional step of calculating the topological charge for each
parameter set. As a result, we obtained good agreement with experiment
with the choice $d_1=1.45$~GeV$^2$, $d_2=0.1$~GeV$^2$ and
$d_3=3.95$~GeV$^2$. Here we choose a quark mass $m(\mu) = 2.6$~MeV defined
at the renormalisation point $\mu^2=170$~GeV$^2$.

In Table~\ref{results}, we present a summary of meson observables
calculated for these two
interactions, with and without the inclusion of the pion
exchange kernel.  One may then compare the effects of unquenching for two
models with dissimilar behaviour in the infrared and intermediate
momentum regions (see Fig.~\ref{fig:coupling}). Though the details of
the deep infrared are generally unimportant for the pattern of dynamical
symmetry breaking in light mesons, we expect a complicated interplay
from the dynamics present in any beyond-the-rainbow truncation.

This difference in the two interactions is seen most prominently for the
pion. Though the effects here are small, as to be expected we see
immediately a difference for the two models. Whilst for Maris-Tandy, the
inclusion of pion-exchange is attractive giving rise to a $2$~MeV shift,
our model exhibits repulsion of the order of $13$~MeV. The leptonic decay
constants see a decrease of 10\% for both models. The effect of
including hadronic contributions is not then \emph{a priori} known, at
least for those mesons most sensitive to dynamical chiral symmetry breaking.

\begin{table}[t!]
\renewcommand{\tabcolsep}{0.5pc} 
\renewcommand{\arraystretch}{1.2} 
\begin{tabular}{@{}c|cc|cc|c}
           & \multicolumn{2}{c|}{Maris-Tandy} & 
	       \multicolumn{2}{c|}{Our Model  } & Experiment\\
           & w/o pi & incl. pi & w/o pi & incl. pi & (PDG~\cite{Yao:2006px})                 \\
	     \hline\hline
$M_\pi$    & $140$ & $138^\dag$  & $125$  & $138^\dag$   & $138$ \\
$f_\pi$    & $104$ & $93.2^\dag$ & $102$  & $93.8^\dag$  & $92.4$\\
           &         & ($90.2$)  &        & ($90.6$)     &       \\
\hline
$M_\sigma$ & $746$ & $598$     & $638$ & $485$      & $400-1200$ \\
$M_\rho$   & $821$ & $720$     & $795$ & $703$      & $776$      \\
$f_\rho$   & $160$ & $167$     & $159$ & $162$      & $156$      \\
           &         & ($167$) &       & ($165$)    &            \\
$M_{a_1}$  & $979$ & $913$     & $941$ & $873$      & $1230$     \\
$M_{b_1}$  & $820$ & $750$     & $879$ & $806$ 	    & $1230$     \\
\hline
$M_{\eta}$ &       &           & $493$ & $497$      & $548$      \\
$M_{\eta'}$&       &           & $949$ & $963$      & $948$      \\
\hline
\end{tabular}
\caption{BSE results for the pseudoscalar, scalar, vector and
axial-vector mesons for the Maris-Tandy and soft-divergence models employed with ('incl.') and without ('w/o') pion backreaction. The parameters marked by $\dag$ are fitted to reproduce the pion mass and decay
constant when the pion-exchange kernel is switched on. Consequently the model parameters,
given in the text, are modified compared to the original Maris-Tandy model
\cite{Maris:1999nt}. Results for the pure rainbow-ladder
kernel without pion back-reaction are for the same parameter set. 
Decay constants given in
parenthesis are those that would result without calculating the full
normalisation condition of the appropriate meson.\label{results}}
\end{table}

For the remaining light mesons, the inclusion of the pion back-reaction
generally leads to a lighter spectrum. In the case of the scalar meson,
we see a reduction of $150$~MeV for both interactions, though the
relative drop is significantly different at 20 and 40\%
for Maris-Tandy and our model respectively.

The vector meson - here identified with the $\rho$ - receives a  negative
shift of $100$~MeV, or $12\%$ for both interactions. In the case of
Maris-Tandy, the resultant rho mass of $\sim 720$ MeV is similar to that
obtained in other studies with this interaction~\cite{Maris:1999nt}. Here,
unquenching has the effect of increasing the decay constant by $5\%$.
Our interaction yields a lighter mass of around $700$~MeV, with
unquenching increasing the decay constant by just $2\%$.  Since our
truncation scheme does not provide a mechanism for the $\rho$ to decay
into pions, we do not observe the tell-tale non-analytic behaviour of 
the rho mass as a function of the pion mass, due to the opening of a decay
channel, as shown in Ref.~\cite{Allton:2005fb}. 

Next, we examine the axial-vectors $a_1$ and $b_1$, which differ by
their charge conjugate properties. As already seen, the net effect of
including the pion back-reaction is a reduction in the meson mass. For
the axial-vectors, we see a $70$~MeV reduction in the bound-state mass
for both charge eigenstates and in both models of the interaction. It is
evident, at least with our tractable unquenching truncation scheme, that
unquenching effects do not resolve the difficulties we have in obtaining
reasonable masses for these mesons. However, we note that the anomalous
mass-splitting in calculations of the $a_1$ and $b_1$ in our model is
half that of Maris-Tandy. In the course of our investigations of
possible parameter sets, we could easily obtain near degenerate
axial-vector masses within our interaction. 

Finally, we take a look at the masses of the $\eta$ and $\eta'$. For our
model, which exhibits a soft-divergence in the gluon momentum, we
calculate a non-zero topological mass following Ref.~\cite{Alkofer:2008et}. 
We see that
the inclusion of the pion back-reaction on the quark propagator has a
minimal effect on the phenomenologically determined mass-splittings.

\section{Conclusions and outlook \label{sec4}} 
Using an approximation scheme for including pion loop effects in
the quark-gluon interaction, we investigated effects of the pion
cloud in the Bethe-Salpeter and Schwinger-Dyson equations. By
considering only the leading amplitude of the exchange pion, moreover
taken from its chiral limit solution, we were able to include the
analytic behaviour beyond the real-value approximation employed in the
previous study \cite{Fischer:2007ze}. The resulting interaction
respects the axial-vector Ward-Takahashi identity.

As a consequence of this technical improvement, we were forced to tackle
the added complication of calculating the full normalisation condition
of the Bethe-Salpeter equation where the momentum exchange of the kernel
need be considered. We found that this indeed gives a sizable
contribution of the order of a few percent that must be properly taken 
into account in order to extract the correct leptonic decay constants.

As a result, we found considerable pion cloud effects in the spectrum of
light mesons. In particular we found a shift of about $100$ MeV
for the mass of the rho-meson in agreement with expectations from chiral
extrapolations of quenched lattice data \cite{leinweber}. 
 
Though this truncation scheme represents a tractable model for including
hadronic unquenching effects, it is clear that spin dependent contributions 
from the Yang-Mills part of the quark-gluon vertex will have a strong 
impact upon the spectrum of light mesons. Such corrections will form a 
part of future investigations.

\section*{Acknowledgements}
We are grateful to Michael Buballa and Dominik Nickel for fruitful discussions 
on the pion back-reaction. We thank Peter Tandy for a critical reading of the
manuscript. This work has been supported by the Helmholtz-University 
Young Investigator Grant No. VH-NG-332.

\begin{appendix}
\section{Pion amplitude at and away from the chiral limit\label{app}}
\begin{figure}[h!]
\centerline{\includegraphics[width=0.9\columnwidth]{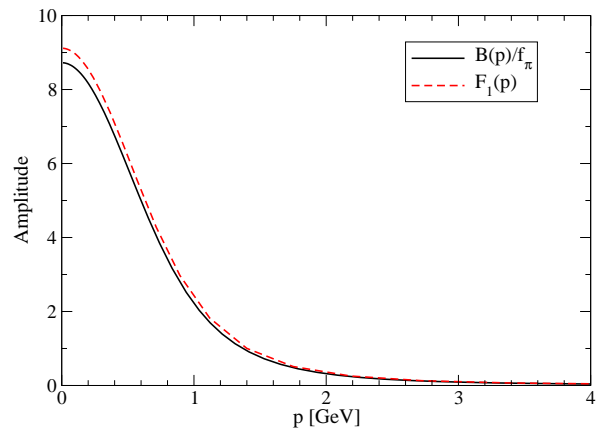}}
\caption{The leading physical pion amplitude $F_1(p,P)$ at
$m_\pi=138$~MeV, compared to the normalised chiral limit value $F_1(p;P)
\approx \frac{B(p^2)}{f_\pi}$. The angular dependence of this amplitude 
is negligible, and chosen here such that $\left(p\cdot P\right)=0$.
}\label{fig:compare}
\end{figure}

One of the key approximations here is that the leading Bethe-Salpeter
amplitude for the pion may be well approximated by the chiral limit
result $B/f_\pi$. It is important for the amplitude to feature the
correct asymptotic behaviour at large momenta, and so for a massive pion
in the exchange kernel we employ the scalar self-energy $B$ as calculated 
in the chiral limit (for complex momenta). The differences between this
and the full amplitude are shown in Fig.~\ref{fig:compare} and are seen
to be at the level of a few percent. 
\end{appendix}


\end{document}